# The Democratic Illusion through the Technological Illusion: a Case Study of the Implementation of a Blockchain to Support an E-voting Platform in Moscow (Active Citizen)

*Hugo Estecahandy,*
*September 2022*

From 1 to 4 July 2022, Muscovites were called upon to vote electronically for the location of the "Luhansk People's Republic" park in the capital. The vote was carried out via the online platform Active Citizen, a project of the Moscow City Hall that allows citizens to vote for or against urban development projects concerning the municipality. In this case, the vote did not ask Muscovites to choose between several names for a particular park, but rather to choose which of three proposed areas of the capital would be given a name that would resonate with current events in the country. Indeed, one of the reasons invoked by Russia to justify the invasion of Ukraine by its army, from 24 February 2022, is the protection and liberation of Donbass, a region in the east of Ukraine, notably composed of the People's Republic of Luhansk and the People's Republic of Donetsk. These Russian-speaking territories, separatists from Ukraine since 2014, were officially recognised by Russia the day after the surprise attack, and are at the center of the Kremlin's war communication to justify its attack, both to the international community and to its own population. Thus, it is a name laden with heavy political symbolism that Muscovites can assign to a piece of their city.

The location of the three proposed territories was indicated by geographical markers, mostly the mention of adjacent streets, and were mostly enriched by specific details, indicated in brackets. Thus, among the nearly 110,000 voters, 56.85% chose the plot "close to the British embassy", while 13.78% opted for a location "between the Lithuanian and Belgian embassies", 10.64% for a territory "close to the German consulate", while 18.73% relied on "the decision of specialists". While voters represented only 5% of Active Citizen registrants (about 2.2 million in total) and less than 1% of the city's total population (about 12.5 million), it is interesting to note that this highly political decision on urban planning was done using an electronic voting platform.

Launched in 2014, Active citizen is one of three e-democracy platforms, a set of digital tools and technologies meant to consolidate the place of citizens in the governance of their territory, set up by the Moscow City Hall. The capital's inhabitants can also express certain problems encountered in their experience of the city on "Moscow - Our city", or propose solutions on a crowdsourcing platform. This use of digital participatory democracy solutions is in line with many projects in other major urban centers around the world that embrace the concept of "e-governance" or "smart city". Furthermore, in 2017, Moscow City Hall announced that this local digital democracy solution would be supported by a "blockchain", a technology underlying cryptocurrencies, which, according to the city hall, makes voting data immutable and freely searchable, as well as the source code, and that every Internet user can verify the integrity of the voting data. The communication states that voting in Active Citizen "is even more open" and that "it will be difficult to say that the administration misinterprets the answers, changes the results of the vote, when citizens themselves can verify this information". Here, the Moscow authorities are playing on a widespread representation, in Russia and around the world, that direct-democratic voting in the Federation is biased and altered by corrupt authorities. As researcher Vladimir Pawlotsky notes, the city is "regularly plagued by corruption cases" (Pawlotsky, 2020a), particularly in the area of urban planning, a sector in which Active Citizen is supposed to strengthen the democratic power of Muscovites. This specialist in urban planning in the Russian capital also explains that when Sergei Sobyanin took over as mayor in 2010, "this unconditional supporter of Russian President Vladimir Putin wanted to improve Moscow's image so that the city could compete with New York, London, Tokyo, Paris and other major capitals. Thus, by the mere implementation, or even mention of the implementation, of a technology, the Moscow City Hall seems to be offering arguments in favor of its desire for transparency and participatory democracy, on the one hand, and its technological advance, on the other.

The question then arises as to whether this implementation of a blockchain to support an online voting solution does not reinforce the political power centralized by the Moscow City Hall more than the distribution of this power among the population. This is also shown by the research of Jacob Zionts (2018) and Caroline Schlaufer (2020), to which this article will return. Similarly, the very use of blockchain to support electronic voting is questioned by some researchers, while explaining that other solutions are more appropriate (Çabuk et al., 2020, Blanchard & al., 2022).

The aim of this article is therefore, firstly, to present the technical contours of Active Citizen and the place of blockchain technology in its operation. Then, we will expose the limits, both technical and ethical, of using a blockchain to support this voting system. Finally, thanks to a lexicometric analysis, we will try to analyze the potential gains for the Moscow City Hall that the use of a blockchain, or rather its announcement effect, entails, especially at the international level.

## I. A very limited technological and democratic contribution

While in 2017 the Moscow City Hall team indicated that the Active Citizen e-voting ecosystem would be supported by a blockchain, the reality is a bit more complex than that. A blockchain is a decentralized digital ledger, which can be replicated in as many copies as there are users, and which binds the recorded data with cryptographic processes that guarantee their inalterability. Created to support the cryptocurrency bitcoin, the blockchain is also characterized by a distribution of authority among all the nodes of the network: all participants theoretically have equivalent rights (to read, modify, decide, etc.). Moreover, the blockchain used within Active Citizen is an adaptation of the Ethereum blockchain, supporting the ether cryptocurrency, promoting the same concepts of decentralization. Thus, without going into the technical details, the use of a voting register that is readable by all and secured, not by an institution to which the population must give its trust, but by mathematical algorithms, could theoretically bring a consolidation of the democratic tool. Following this logic, the assertion of researcher Leonid Smorgunov is acceptable. He noted in 2018 that "in the theory of public policy, there is an eternal gap of fair procedures that could be used to achieve a fair outcome in the decision. Blockchain, as a network of distributed registers, often positions itself as an institution ensuring fairness in decisions by voting based on a consensus procedure" (Smorgunov, 2018.)

Since the first blockchain in 2008, and especially since the technology started to be industrially adapted in 2015/2016, derivatives of the original concept have been created. Thus, it was the appearance of distributed ledger technologies (DLT), proposing blockchains with a distribution of authority no longer distributed but shared, where only a few nodes of the network formed by the copies of the ledger had the maximum level of authority, or centralized in a single node that has authority over the others.

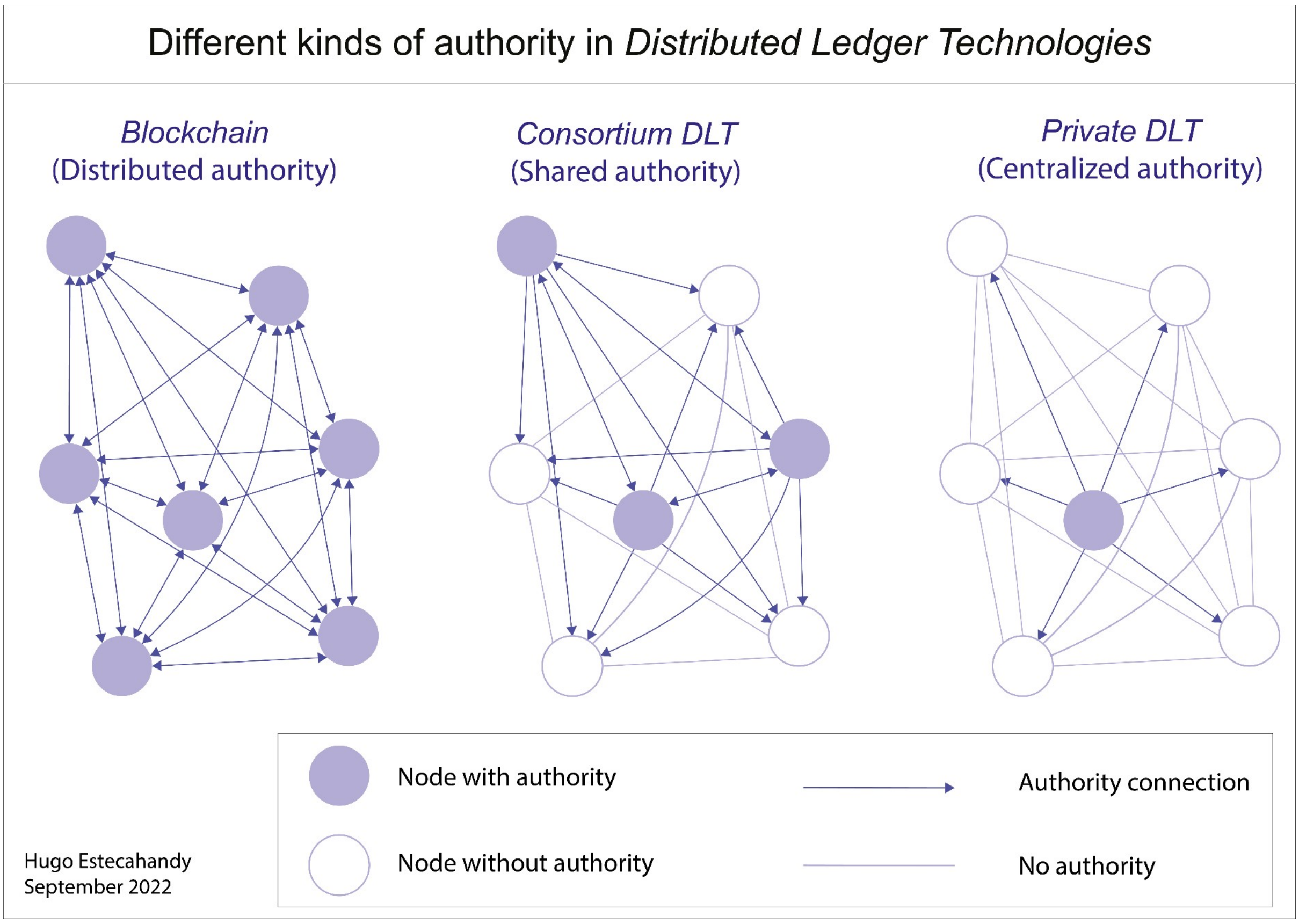

*Figure 1*

In fact, the blockchain that supports online voting on Active Citizen seems to belong to the last of these categories. Indeed, if all the nodes of the network can read the data recorded on the register, there is an entity, in this case the Moscow City Hall - or rather the Internet or mobile application of Active Citizen via which the Internet users vote - which can write information, in this case the vote, its timestamp, the secret identifier of the voter. There is thus an important link in the Moscow City Hall's implementation of a blockchain, since the voting information is written on a node (Master node) that it controls, from which the other nodes are updated and can check that no information is subsequently altered.

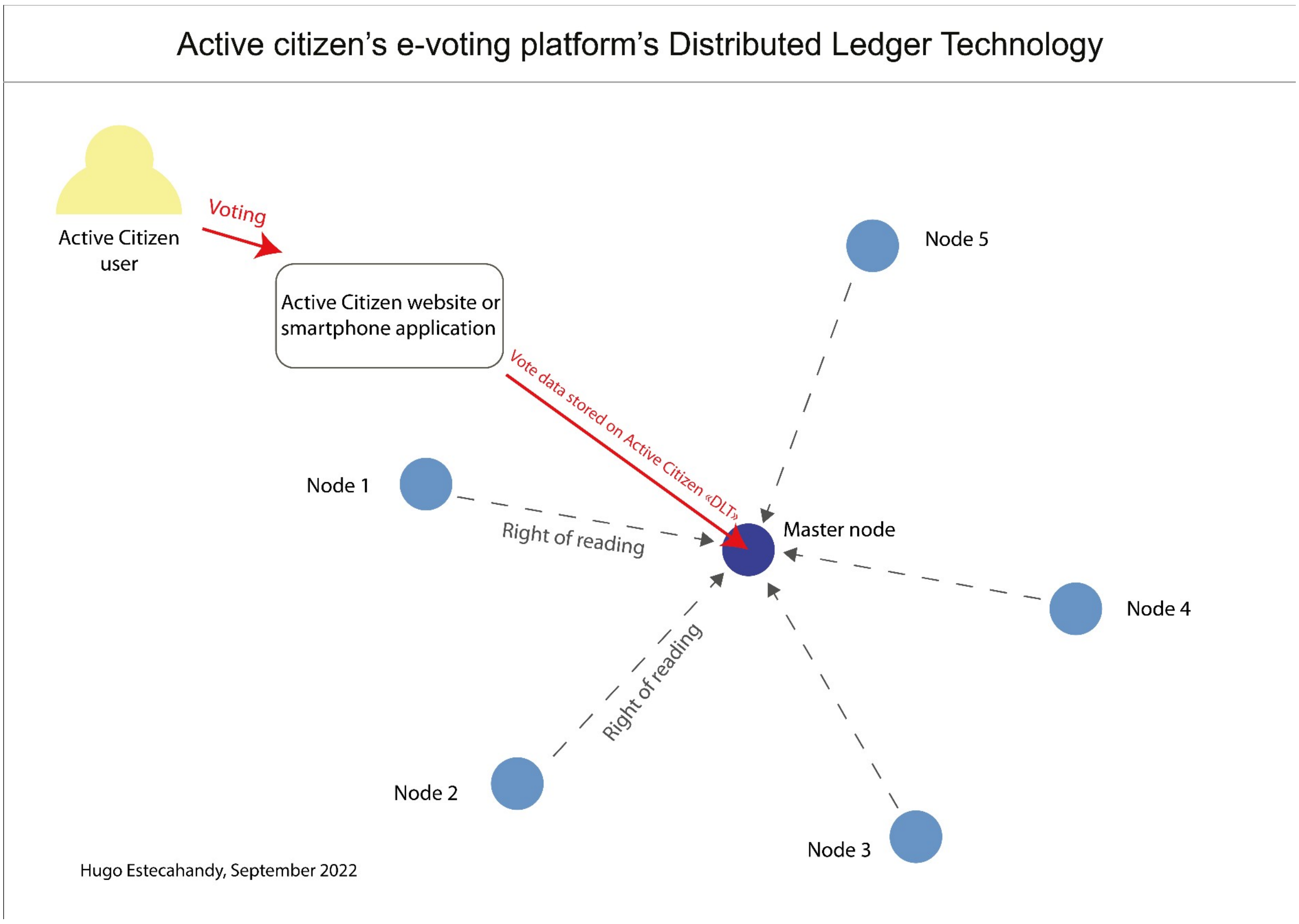

*Figure 2*

Active Citizen's DTL ultimately allows some observers to ensure that no information in the registry is altered, but the validity of this information is the responsibility of one actor: Moscow City Hall. As Blanchard et al. (2022) write, "the link between the real world and the blockchain depends on the fact that all parties commit to rules. The only truth that is guaranteed by writing information on a blockchain is that the information in question is written on the blockchain in question. There is no evidence here that the data originally written is reliable.

## II. Active citizen or the democratic illusion

Thus, as we have just seen, the implementation of a blockchain to support the Active Citizen voting system is only very limited, both technologically and democratically. However, Moscow City Hall has an interest in offering this service to its citizens, whether it is secured by a blockchain or not. Researcher Caroline Schlaufer has looked at the case of Active Citizen, asking "Why do non-democratic regimes promote e-participation" (Schlaufer, 2020). She argues that simply offering an online voting system, "can accommodate different legitimation strategies into one single online tool. A well-designed e-participation tool serves, first, as an input channel for citizen preferences, thereby enhancing input legitimation. Second, it can increase output-based legitimation by increasing citizen satisfaction with policy performance. Third, an e-participation platform also disseminates the narrative of a modern, inclusive and responsive government. Thus, Active Citizen would above all give Muscovites the illusion that they are integrated into a more democratic process thought out by an administration at the cutting edge of new technologies.

This democratic illusion is even more perceptible when we look at the types of votes that are proposed. For example, between January 2019 and mid-April of the same year, the five projects with the most votes were the opening of a pediatric operating theater in a city hospital, a 24-hour helicopter medical evacuation service, the creation of a "protected nature zone" in the Bratayevskaya Poima Park to preserve local wildlife, free admission to the city's museums once every three weeks and, finally, the introduction of modern trains to connect the Moscow Belarusskaya station with the city of Odintsovo (Moscow Oblast). Moreover, as Jacob Zionts writes, "Active Citizen does little to change the underlying distribution of power in Moscow politics: voting on the name of a new metro line does not disrupt endemic corruption; selecting new seat colors for the Luzhniki Big Sports Arena will not empower investigative journalists. Active Citizen grants Muscovites limited self-expression but only in a safe and sanitized form's (Zionts, 2018).

The researcher draws a parallel between the use of the platform and the legitimization of urban projects that could have led to public rejection. In 2017, with the blessing of Russian President Vladimir Putin, Moscow Mayor Sergei Sobyanin planned the destruction of several thousand homes: Soviet buildings that he wanted to replace with modern real estate projects by wealthy developers. As anger began to grow among the population of the capital, a "referendum" was proposed to the inhabitants via Active citizens. They could vote for or against the 'renovation' of their buildings, to which 90% of them replied in the affirmative. Then, "officials quickly pointed to the overwhelming support among residents as support for the demolition." As one of the interviewed people said to Schlaufer during her research, 'of course, if a majority of citizens agreed with Active Citizen about something, they will, afterwards, not complain about it. And it leads to a higher satisfaction of the city government's activities' (Schlaufer, 2020). In addition to giving a democratic illusion to some Muscovites, the voting platform would allow the mayor's office to make them accept, in spite of themselves, policies that they would logically have opposed.

## III. An international digital showcase for Moscow?

But this use of an electronic voting platform does not only seem to allow the Moscow mayor's office to consolidate its power and its image in Moscow or in Russia, but also internationally. As Vladimir Pawlotsky explains, Mayor Sobyanin's aim is to make Moscow "a global city" and "a world center of financial stability and innovation" that "an economy dependent on the - very volatile - price of hydrocarbons" requires. Moscow sought - since the war in Ukraine has surely put this strategy into question - 'as a privileged platform for foreign investment, to capture international economic and financial flows by hosting the headquarters and subsidiaries of transnational corporations, developing cultural activities as well as tourism' (Pawlotsky, 2020a). Thus, Moscow, and by extension Russia, are in a logic of city branding and nation branding internationally. The 2018 Football World Cup helped to present the city and the rest of the country to the world.

The Active Citizen project and the announcement of the implementation of a blockchain also meet the objectives of this strategy, communicating a democratic and innovative image of Moscow to the rest of the world. Thus, the announcement of the addition of a blockchain to Active citizen was shared internationally, by more or less specialized English-speaking media. It is interesting to note that most of these articles repeat the December 2017 press release from the city hall and its language. Let's remember that this communiqué made much of the codes dear to cryptoanarchists and other defenders of a free Internet, some of which are founding principles of blockchain: distribution of the verification of the voting process, free access to the underlying protocol or incorruptibility of the decisions taken by Muscovites concerning the future of their city.

In order to test our hypothesis that the foreign media treatment of the implementation of a blockchain within Active Citizen has helped to project a democratic and modern image of Moscow, we conducted a lexicometric analysis on 6 of these English-language press articles (see Appendix 1 for the article references). This analysis has been conducted through the software IRaMuTeQ, developed at Toulouse's University of Mirail (France). Above all, the software allows us to highlight classes of words that it will automatically detect using a classification method called the Reinert method. As it can be seen on the Figure 3, the Reinert classification brings out 5 classes of words, out of the text corpus formed by our 6 articles.

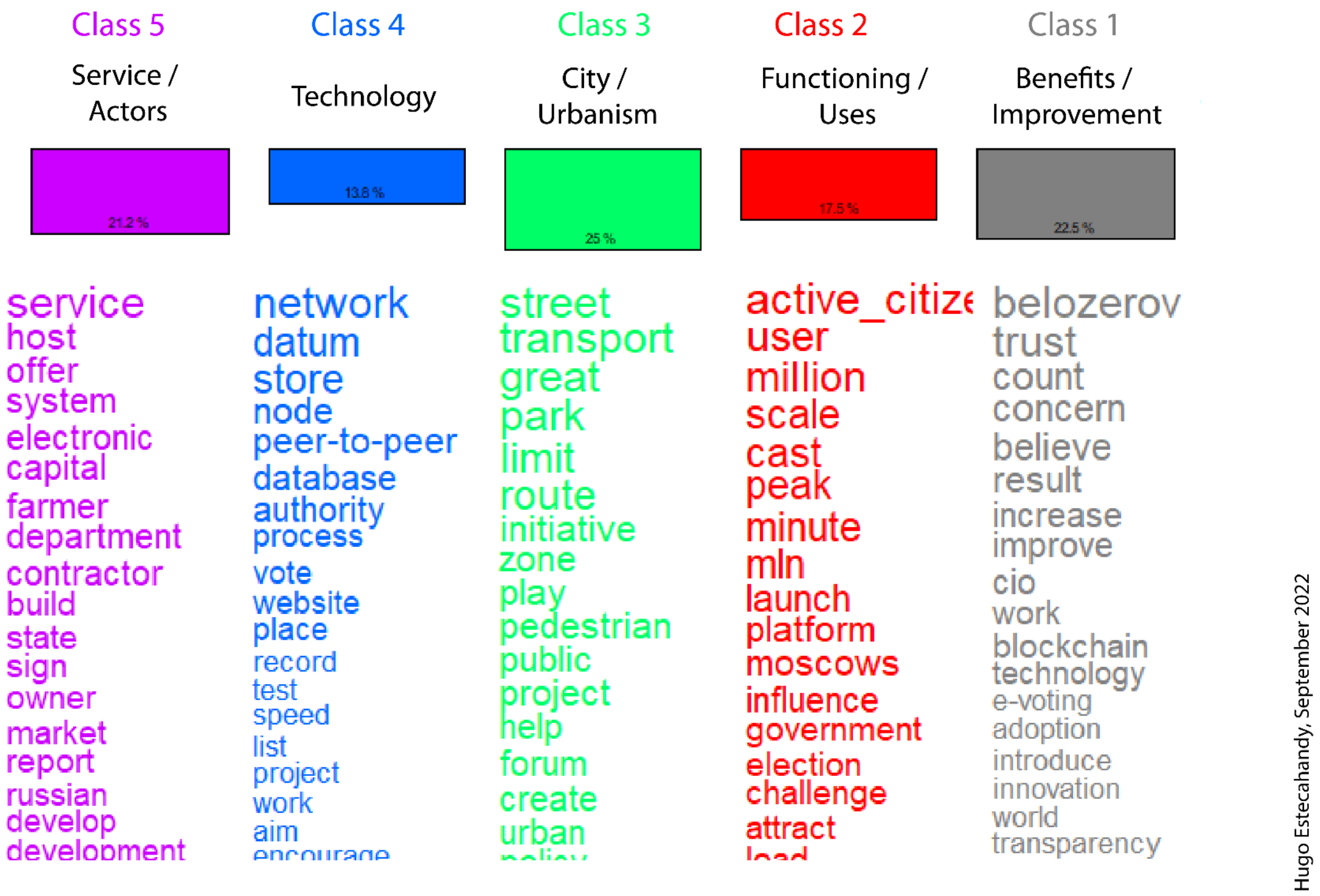

*Figure 3*

Class 1 is the most interesting for us, as it refers to all the words that refer to the positive impact of the implementation of blockchain by the Moscow City Hall. And among these words we find several terms that are part of the lexical field of blockchain technology: "trust," "transparency," "innovation," "improve," "increase."
Then we also performed a similarity analysis on the same body of text. This graph shows the words and their frequency. The larger the word, the more often it appears in the text. The words are spatialized according to their proximity to the text.

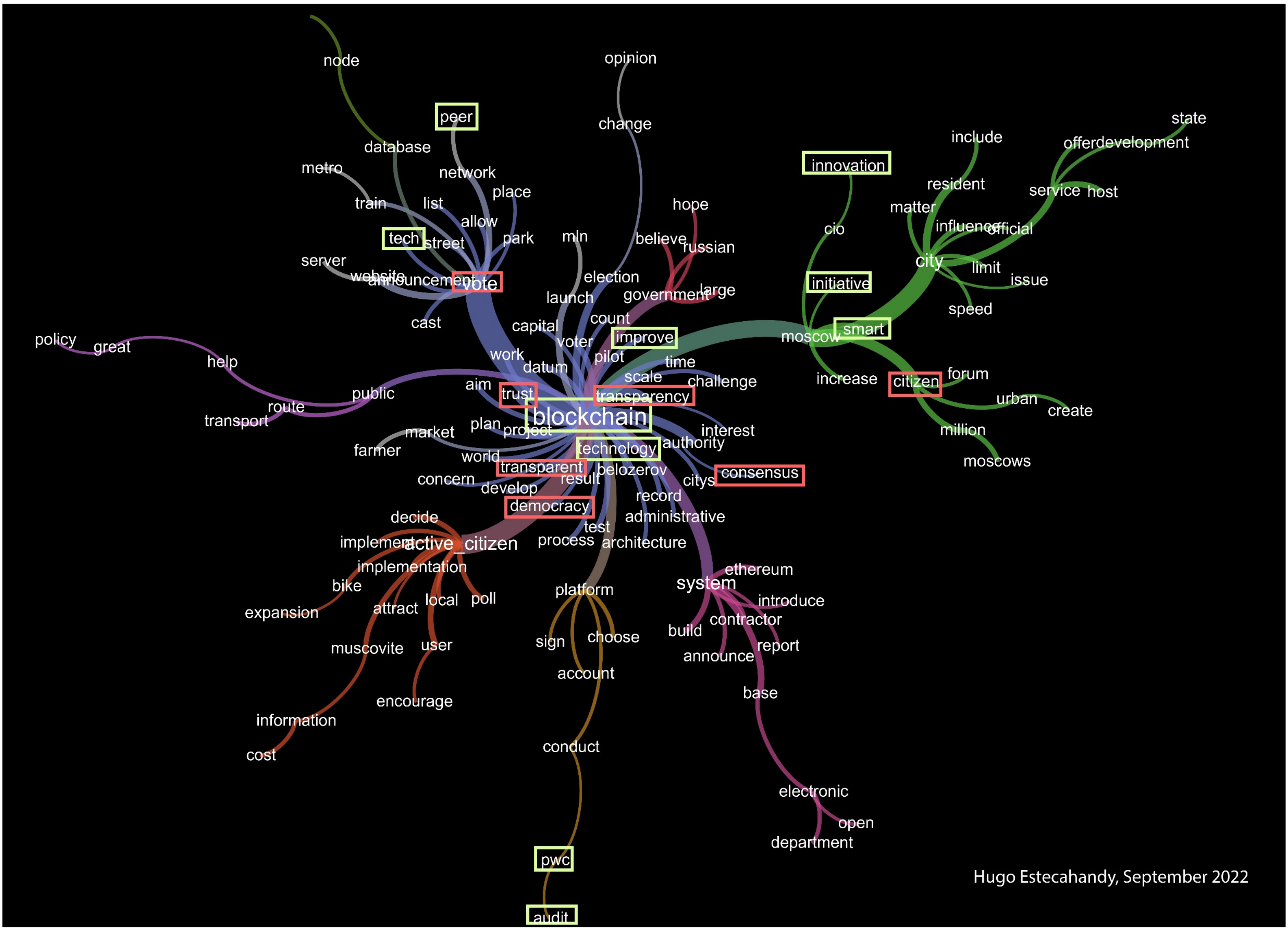

Figure 4

Blockchain is the most present word as we have seen on the cloud and it is in the center of this graph. What we see here is interesting for two reasons: many words in the lexical field of democracy (in red) are directly related to the term blockchain, a term to which the words "Moscow" and "Activie_Citzen" are also strongly related. The lexical field of innovation is also very present, with terms directly linked to "Moscow." This graph allows us to see how the media’s treatment of the implementation of a blockchain by the Moscow City Council served its international communication objectives.

## Conclusion

The implementation of a blockchain in the voting system of Active Citizen has above all allowed to consolidate the discourse that the Moscow City Hall wishes to attach to this application. Thus, with the Active cities project "the idea is not to make Muscovites vote on major issues - far from it - but to ask them questions designed to elicit answers that correspond to the objectives of the city government, thus helping to legitimize its policies (Pawlotsky, 2020b). From what we have seen, these goals were also a consolidation of the positive feelings of its population towards a democratic and modern city. It also participated in the creation of an international brand image of Moscow as we could understand through the analysis of press materials of foreign media. Recently, we could observe what could be a new type of recourse to Active Citizen for the authorities, which allowed renaming a part of the urban area in front of the British embassy and thus artificially correlate what would represent the ideas of Muscovites with the geopolitical ambitions of the Kremlin and its discourse of justification of the war in Ukraine. In addition to the above-mentioned benefits, Active Citizen seems to allow the Moscow City Hall to attract even more favors from the Kremlin under the cover of a completely illusory participatory democracy. And the announcement of the implementation of blockchain has contributed to this.

Through the common thought that "blockchain technology provides the kind of transparency and immutability that could sustain democracy, Moscow's implementation of the technology through Active Citizen has provided only the illusion of empowerment; the political system in Moscow remains autocratic" (Zionts, 2018). And this allows us to go further and quote Schlaufer as our results also showed "that the introduction of e-participation in a non-democratic regime is not a contradiction. Therefore, controlled e-participation is not a challenge to the underlying distribution of power between governments and citizens" (Schlaufer, 2020).

**Annexe 1 (Analyzed articles references) :**

1. « Moscow introduces blockchain e-voting », *SmartCitiesWorld*, 12/05/2017.

2. « Moscow to Develop a Blockchain System for Transparent City Services », *CoinDesk*, 08/15/2019.

3. « Moscow to pilot blockchain platform for e-voting », *FinExtra*, 12/04/2017.

4. « Russia Is Leading the Push for Blockchain Democracy », *CoinDesk*, 02/18/2018.

5. « How Moscow is pioneering the rise of smart cities », *CIO,* 11/06/2017.

6. « This is How Russia Utilizing Blockchain to Enhance Democracy », *Sputnik News*, 03/02/2018.